\documentclass{acm_proc_article-sp}

\usepackage[utf8]{inputenc}
\usepackage{graphicx}
\usepackage{grffile}
\usepackage[pdftex]{hyperref}
\usepackage{booktabs}
\usepackage{subfigure}

\begin{document}

\title{
  Social Networking by Proxy:  A Case Study of \\ Catster, Dogster and Hamsterster 
}

\numberofauthors{2}

\author{
  Daniel Dünker \\
  \affaddr{University of Koblenz--Landau} \\
  \affaddr{Universitätsstr.\ 1, 56070 Koblenz} \\
  \email{dduenker@uni-koblenz.de}
  \and
  Jérôme Kunegis \\
  \affaddr{University of Koblenz--Landau} \\
  \affaddr{Universitätsstr.\ 1, 56070 Koblenz} \\
  \email{kunegis@uni-koblenz.de}
}

\maketitle

\begin{abstract}
  The proliferation of online social networks in the last decade has not
  stopped short of pets, and many different online platforms now exist
  catering to owners of various pets such as cats and dogs.  These
  online pet social networks provide a unique opportunity to study an
  online social network in which a single user manages multiple user
  profiles, i.e.\ one for each pet they own.  These types of
  \emph{multi-profile networks} allow us to investigate two questions:
  (1) What is the relationship between the pet-level and human-level
  network, and (2) what is the relationship between friendship links and
  family ties?  Concretely, we study the online pet social networks
  Catster, Dogster and Hamsterster, the first two of which are the two
  largest online pet networks in existence.  We show how the networks on
  the two levels interact, and perform experiments to find out whether
  knowledge about friendships on a profile-level alone can be used to
  predict which users are behind which profile.  In order to do so, we
  introduce the concept of multi-profile social network, extend a
  previously defined spectral test of diagonality to multi-profile
  networks, define two new homophily measures for multi-profile social
  networks, perform a two-level social network analysis, and present an
  algorithm for predicting whether two profiles were created by the same
  user.  As a result, we are able to predict with very high precision
  whether two profiles were created by a same user.  Our work is thus
  relevant for the analysis of other online communities in which users
  may use multiple profiles. 
\end{abstract}

\begin{figure}
  \centering
  \includegraphics[width=1.00\columnwidth]{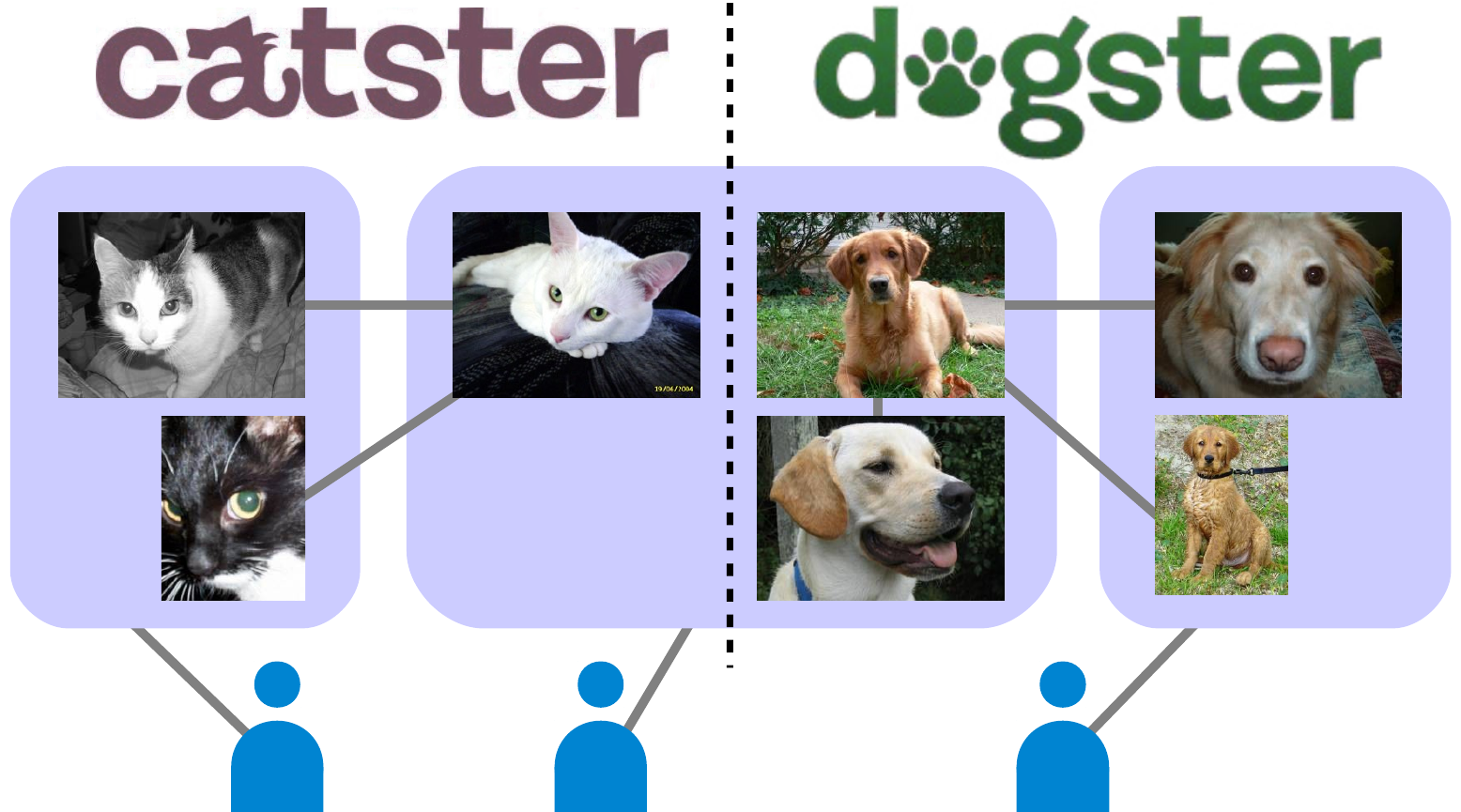}
  \caption{
    \label{fig:er}
    Entities and relationships in Catster and Dogster. 
    Individual owners may own both cats and dogs, but
    friendship links across the two sites are not possible.
    The sites are both unconnected to Hamsterster. 
  }
\end{figure}

\section{Introduction}
Pet ownership is common in many countries.  In the United States for
instance, 47\% of households owned at least one dog, and 46\% at least
one cat in 2012 \cite{appa}.  It therefore comes as no surprise that
specialized online social networking platforms exist specifically for
pets.  In general, online social networks may range from the very
generic such as Facebook and Twitter, to the very specialized for
dedicated communities related to hobbies, activities or professions.
Nevertheless, the specific topic that unifies the community usually does
not affect the basic mechanism of an online social network: A user
creates an account to connect with other users.  Online pet social
networks are however different in this regard. In online pet networks,
users can create any number of accounts, one for each pet they own.
While individual persons cannot usually be stopped from creating
multiple accounts in an ordinary online social network, this is usually
frowned upon.  On Wikipedia for instance, the use of multiple accounts
by a single user is restricted to a narrow list of special cases which
includes testing, running bots, or users which have been assigned
official roles.  Outside of these, the use of multiple accounts is
proscribed, and when used for disruption is called \emph{sock puppetry}
\cite{sockpuppet}.  For these reasons, information on the use of
multiple accounts by users in online social networks is a rarely studied
problem, and few datasets for its study exist.  As an example, one
study performs the task of predicting whether a given Wikipedia account
is a sock 
puppet on less than a hundred accounts \cite{solorio2013}.  In contrast
to this, we are able to perform a study on several hundreds of thousands
of users in this paper.

\begin{table*}
  \caption{
    \label{tab:species}
    Datasets analysed. 
  }
  \centering
  \begin{tabular}{ l r r r r }
    \toprule
    \textbf{Dataset} & \textbf{\#Pets} & \textbf{\#Friendships} & \textbf{\#Households} & \textbf{Pets per household} \\
    \midrule
    Catster           & 204,424 &  5,443,885 & 105,089 & 1.95 \\
    Dogster           & 451,710 &  8,543,549 & 260,390 & 1.73 \\
    Catster + Dogster & 623,766 & 13,991,746 & 333,111 & 1.87 \\
    Hamsterster       &   2,950 &     12,531 &   1,575 & 1.87 \\
    \bottomrule
  \end{tabular}
\end{table*}

In online pet social networks, a single user may (and is expected to) create
one account for each owned pet.  All social networking functionality
such as entering personal information, creating friendship links to
others, etc., are then performed on the pet level.  
Figure~\ref{fig:er} illustrates how the multiple pet profiles created by
a user form a family of pets. 
With their structure that allows multiple profiles per account, 
online pet social networks thus make it possible to investigate the following 
questions:
\begin{itemize}
  \item How does the fact that individual users own multiple profiles
    influence the structure of the social network?
  \item Is it possible to predict that two accounts are managed by the same
    person? 
\end{itemize}

These questions are analysed under multiple aspects in the remainder of
the paper. 
In Section~\ref{sec:related}, we review related work and in Section
\ref{sec:dataset}, we describe our three datasets.  
In Section~\ref{sec:network}, we perform social network analysis,
  in order to determine crucial differences between both networks. 
In Section~\ref{sec:homophily}, we investigate the homophily on
  both levels, asking whether the account-level network is characterized
  by higher homophily values, and if yes, for which node properties this
  is true. 
In Section~\ref{sec:spectral}, we perform a spectral analysis of
  the networks, for which we introduce an extended spectral diagonality
  test in order to compare friendships with family ties. 
In Section~\ref{sec:prediction}, we analyse the problem of
  predicting that two profiles were set up by the same account, with the
  goal to find out whether this is possible at all, and if yes which
  structural and metadata properties are suitable for this task.
Section~\ref{sec:conclusion} concludes the paper. 

\section{Online Pet Networks}
\label{sec:related}
The analysis of social networks has its roots in the social
sciences~\cite{b268}.  More recently, the use of social network datasets
extracted from online social networking platforms have led to a large
amount of research in computer science and network science.  Online
social networks allow people to connect via a platform in order to
communicate, share content, or simply manage a list of connections for
various purposes.  In most such platforms, a single user account is used
to manage a single user profile, to which users can add information such
as their age, location, sex, favorite movies, songs, food, or any other
metadata deemed interesting to the particular community.  In only few
cases can multiple profiles be created by a single user.  An example is
given by company or product pages on Facebook, of which one user can
create more than one.  In that case however, there may be more than one
user managing each profile, resulting in group-like semantics rather
than profile-like semantics.  In most online social networking platforms,
the creation of multiple profiles by one user is not allowed, only
possible by using multiple email addresses, or restricted to very
specific users. On Wikipedia for instance, multiple accounts created by
a single person are referred to as \emph{sock puppets}, and are
proscribed \cite{sockpuppet}.  Therefore, few datasets are available and
only little research has been
conducted on the topic, an example being the detection of sock puppets
on Wikipedia \cite{solorio2013} using one hundred accounts.  Text mining
approaches to detect sock puppets in Wikipedia have been described, too
\cite{solorio2013b}. Therefore, online pet social networks such as
Catster, Dogster and Hamsterster present a unique opportunity to study a
social network in which users manage multiple profiles.  What is more,
due to the fact that this is not proscribed by the sites, but instead
represents the normal way of using them, information about identical users
is openly available on these sites, making this study possible.

Many specialized online networking platforms exist,
and online pet social networking platforms specifically have been
studied before, 
although social network analyses have not been performed on them.
Related   
work analysing online animal social networks has covered Catster,
Dogster and Hamsterster, but only used small samples of the full
networks for analysis:  2,000 dogs and 2,000 cats in \cite{golbeck2009},
and 10,000 dogs and 10,000 cats in \cite{b590}.  
None of these works performs a network analysis.  The latter
paper asks the question whether knowledge about family
ties can improve prediction of friendship ties; the question
is answered positively. 

%
%
%

A distinct topic is that of animal networks such as networks of sheep
\cite{konect:hass}, dolphins \cite{konect:dolphins} and macaques
\cite{konect:takahata1991}.  Those refer to social networks in which the
actors are (usually wild) animals, whose social ties are not conditioned
by humans. 
Another distinct concept is that of circles, as used for instance in
Google+ \cite{b856}.  Although families in pet networks have been called
\emph{circles} (e.g.\ in \cite{b590}), they are not the same concept as
used on Google+.  On Google+, a circle is a device to group one's own
friends. Hence, circles do not provide a new type of link beyond
friendships, and cannot be compared to the families of online pet social
networks. 

\begin{figure*}
  \centering
  \subfigure[Distribution of join dates]{
    \label{fig:joined}
    \includegraphics[width=0.90\columnwidth]{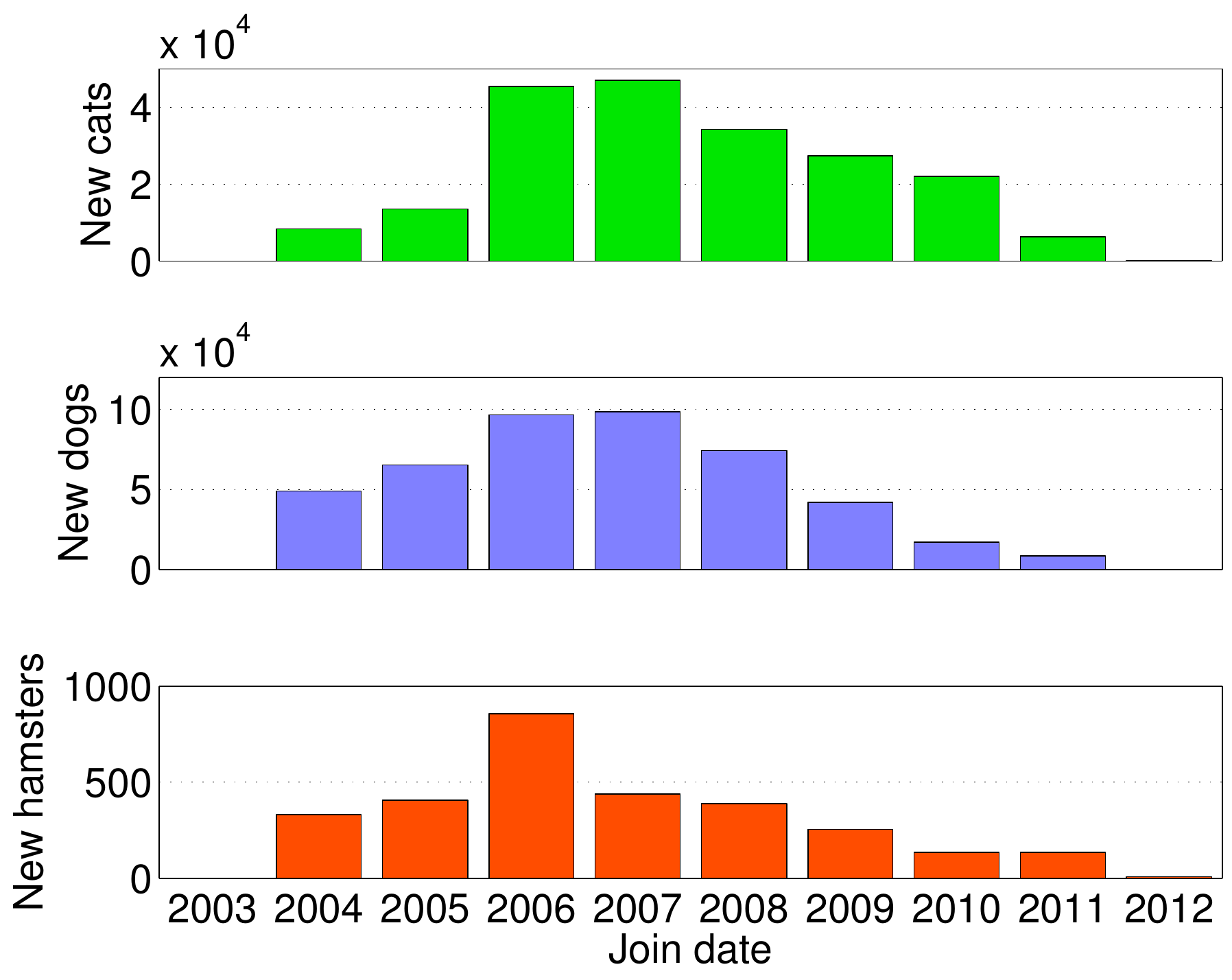}
  }
  \subfigure[Population pyramid of Catster]{
    \label{fig:pyramid-cat}
    \includegraphics[width=0.55\columnwidth]{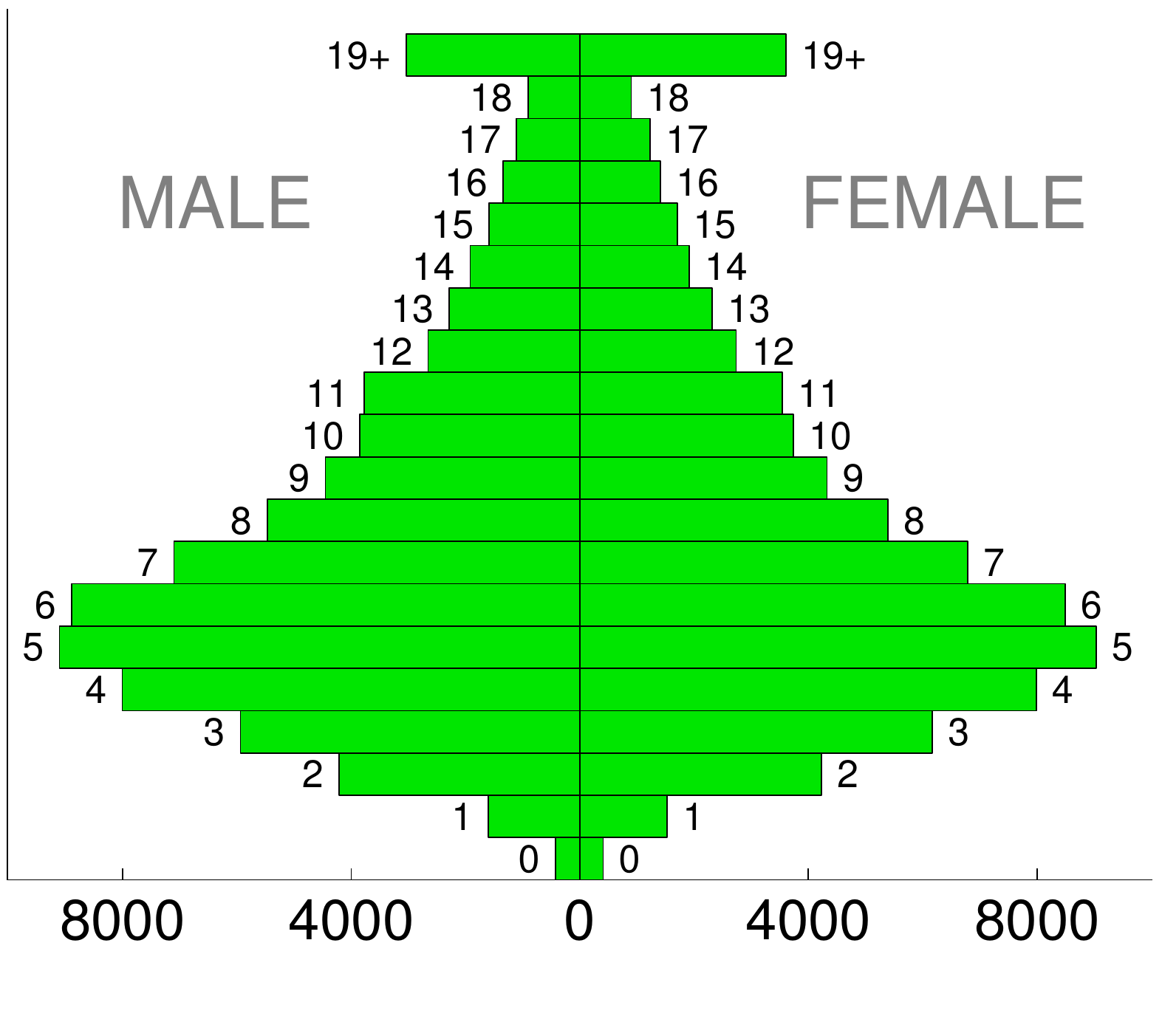}
  }
  \subfigure[Population pyramid of Dogster]{
    \label{fig:pyramid-dog}
    \includegraphics[width=0.55\columnwidth]{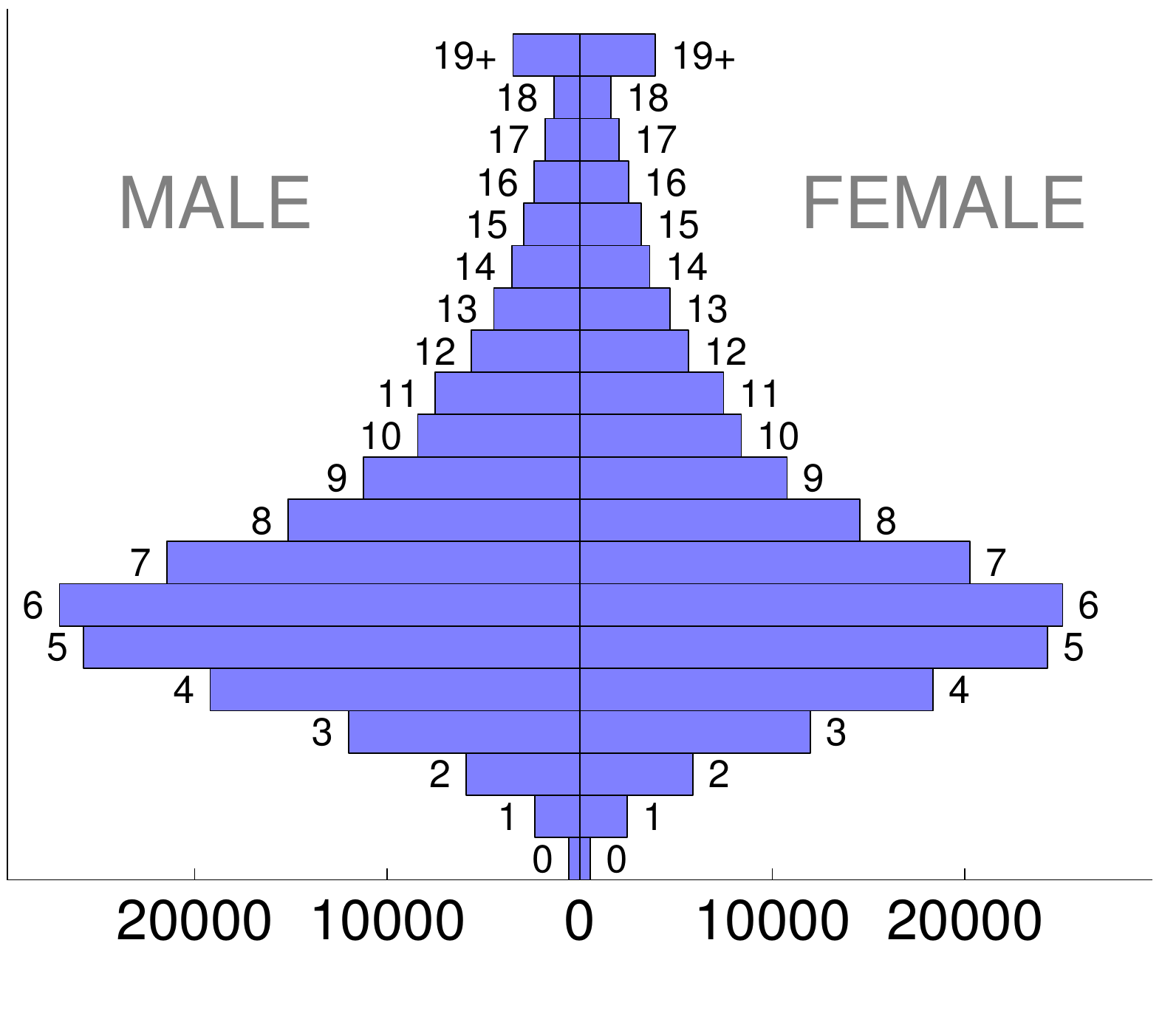}
  }
  \caption{
    \label{fig:demography}
    Demographic characteristics of the pet networks. 
    (a)~Distribution of join dates, i.e., pet profile creation dates. 
    The oldest profiles have dates in 2003, on Dogster and Hamsterster.
    The newest accounts crawled by us were created in early 2012.  
    (b-c) Population pyramids of Catster and Dogster, showing the
    distribution of ages and sexes.  
  }
\end{figure*}

\section{Datasets}
\label{sec:dataset}
We use datasets of Catster, Dogster and Hamsterster.  Since Dogster and
Catster share user accounts, we also report statistics on
the union of these two. 
An overview of the datasets is given in Table~\ref{tab:species}. 
\texttt{catster.com} and \texttt{dogster.com} were both founded in
2004~\cite{founded}. 
Both sites are 
linked:  A single user can create pet profiles on both sites, and
individual cat and dog profile pages are interlinked via a family link
when they were created by the same user.  \texttt{hamsterster.com} is an
independent 
site created in 2003 or 2004.\footnote{The exact creation date of
  Hamsterster is not known to us.  The oldest accounts there date from
  2003, but the domain \texttt{hamsterster.com} was registered in
  2004 \cite{hamsterster-domain}, and the phrase ``after nearly ten
  years'' written in October 2014 on Twitter\textsuperscript{2} suggests
  a creation date of 2004.}  
Hamsterster appears to have
been shut down as of October 2014.\footnote{As of October 2014, the
  Twitter account @HAMSTERster™ 
states that Hamsterster had been closed ``after nearly ten years''. 
}
Other such ``online social petworks'' exist, such as   
\texttt{bunspace.com} for rabbits, but are not studied in this paper.
The suffix \emph{-ster} in these names was likely chosen as a reference to
\texttt{friendster.com}, created in 2002.  
We crawled Catster and Dogster from August 2011 to March 2012, 
and Hamsterster in February 2012. 

On all three sites, a single user can create accounts for any number of
pets.  Catster and Dogster are connected, and thus a single user account
can be used for both sites, although 90.3\% of accounts across Catster and
Dogster include only cats or only dogs.  The group of pet profiles created by a
single user makes up a \emph{household} or \emph{family}. 
Friendship links are allowed within a single household in Dogster and
Catster, but are not allowed in Hamsterster. 
All friendship links are undirected. 

Catster and Dogster allow only cats (\emph{Felis catus}) and dogs
(\emph{Canis lupus familiaris} or \emph{Canis familiaris})
respectively.  Hamsterster allows multiple species of hamsters
(subfamily Cricetinae) and gerbils (subfamily Gerbillinae), the most
common species being the golden hamster (\emph{Mesocricetus auratus}).  
The Hamsterster dataset contains at least one cat, a rat
and five guinea pigs.
We also found profiles in all three platforms
apparently created for multiple pets (e.g., named ``Hamster babies''). 
For each of the three sites, about two thirds of all users are located in the
United States.

\section{Multi-profile Social Network \\ Analysis} 
\label{sec:network}
The multi-profile social networks of Catster, Dogster and Hamsterster
can be analysed using tools of social network analysis on two different
levels:  the profile level (pet level
and the account level (family or household level).  By
performing
social network analysis, we can derive 
several properties from a multi-profile social network.  First, we can
derive the differences and similarities between the two networks.
Second, we can ask which of the two is more similar to a typical social
network, in order to assess whether the network is better modeled as an
account-level network to which profiles are attached, or a profile-level
network in which the profiles are aggregated into groups. 

\subsection{Definitions}
We now introduce a formal definition of a
\emph{multi-profile social network}, of which Catster, Dogster and
Hamsterster are examples.  
A multi-profile social network is a social network in which each person
is associated with one or more profiles, and in which the actual social
relationships as well as the metadata such as age, sex and location are
associated to individual profiles.  In the online case, a multi-profile
social network allows each user to manage one or more
profiles.  The set of profiles managed by a single account in a
multi-profile social network may also be called a \emph{household} or a
\emph{family}.  The latter term in particular is used by the three
studied online pet social networking sites. 

We denote a multi-profile social network by $G=(V,W,E,m)$, where $V$ is
the set of profiles, $W$ is the set of accounts, $E \subseteq V \times
V$ is the set of friendship edges connecting profiles, and $m: V
\rightarrow W$ is a mapping from profiles to accounts.  
Individual profiles will be denoted by the letters $u$, $v$, etc., while
accounts will be denoted by the letters $i$, $j$, etc. 
As in other social networks, additional metadata for profiles, accounts and
friendships may be defined.  The online pet social networks we study
include extensive profile metadata (described in
Section~\ref{sec:homophily:methodology}), but do not include account
metadata, because they present everything from the point of view of the
pet.  The graph $G_{\mathrm p} = (V,E)$ then represents the profile-level social
network, while $G_{\mathrm a} = (W, m(E))$ represents the
account-level social network, using the definition
\begin{align}
  m(E) = \{ & \{i,j\} \mid i \neq j \wedge \exists\;\{u, v\} \in E :
  \nonumber \\
  & m(u) = i \wedge m(v) = j \},
\end{align}
that is, $G_{\mathrm a}$ is the result of identifying vertices in
$G_{\mathrm p}$ that are in the same household, not including loops in
  the result. 
An overview of the differences between the two levels of networks is
shown in Table~\ref{tab:statistic} in terms of numerical statistics. 

\begin{table*}
  \caption{
    \label{tab:statistic}
    Network statistics in the profile-level and in the account-level social
    network for the three sites. 
  }
  \centering
  \begin{tabular}{ l | r r r | r r r }
    \toprule 
    & \multicolumn{3}{|c|}{\textbf{Profile-level network}} & 
    \multicolumn{3}{|c}{\textbf{Account-level network}} \\
    & \multicolumn{3}{|c|}{$G_{\mathrm p} = (V,E)$} & 
      \multicolumn{3}{|c}{$G_{\mathrm a} = (W, m(E))$} \\
    \textbf{Statistic} & \textbf{Cat} & \textbf{Dog} & \textbf{Ham.} &
    \textbf{Cat} & \textbf{Dog} & \textbf{Ham.} \\ 
    \midrule


    \#Nodes & 204,473 & 451,710 & 2,952 & 105,138 & 260,390 & 1,576 \\
    \#Edges & 5,448,197 & 8,543,549 & 12,534 & 494,858 & 2,148,179 & 4,032  \\
    Average degree & 53.29 & 37.82 & 8.49 & 9.41 & 16.50 & 5.12 \\
    Largest connected component & 72.79\% & 94.42\% & 60.57\% & 64.98\% & 98.30\% & 55.46\% \\
    Power-law exponent\textsuperscript{a} & 
    2.12 (19) & 2.15 (26) & 2.46 (20) & 2.27 (8) & 2.27 (18) & 2.14 (7) \\
    Gini coefficient\textsuperscript{b} & 77.10\% & 75.06\% & 61.06\% & 72.93\% & 72.36\% & 63.02\% \\
    Clustering coefficient & 1.10\% & 1.43\% & 9.04\% & 0.38\% & 1.01\% & 13.13\% \\
    Diameter\textsuperscript{c} &
    10 & 11 & 14 & 10 & 10 & 8 \\
    Mean path length\textsuperscript{c} &
    2.73 & 3.39 & 3.42 & 2.62 & 3.36 & 3.17 \\
    \bottomrule
    \multicolumn{7}{l}{
      \textsuperscript{a} The minimum degree $d_{\min}$ at which the power law was
      fitted is shown in parentheses
    } \\
    \multicolumn{7}{l}{
      \textsuperscript{b} Measured using the method from \cite{kunegis:power-law}
    } \\
    \multicolumn{7}{l}{
      \textsuperscript{c} Measured in the largest connected component 
    } 
  \end{tabular}
\end{table*}

\subsection{Demographic Characteristics}
The distribution of sexes and and ages of pets is shown in 
Figure~\ref{fig:demography} (b-c). Both sexes are equally distributed in Catster
and Dogster, and the age distribution reflects the pet's life spans. 
On average, there are two pets to one household.
The average number of pets per household is consistent over all three
pet types; it is 1.95 for cats, 1.73 for dogs and 1.87 for hamsters 
(see Table~\ref{tab:species}).  
The distribution of pets per household (shown in Figure~\ref{fig:pph})
is power law-like, with similar power law exponents for all three
sites.  
The fitted power law exponents using the method described in
\cite[Eq.\ (5)-(6)]{b408} are 3.62 for Hamsterster ($p_{\min} = 5$), 3.63 for
Catster ($p_{\min} = 6$), 3.90 for Dogster ($p_{\min} = 4$) and 3.79 for
Catster and Dogster combined ($p_{\min} = 5$).  The fitted parameter
$p_{\min}$ denotes the starting point of the fit.

The fact that the number of pets per household follows a power-law
distribution closely is interesting.  In usual social networks, this is
explained through a process of preferential attachment, i.e., persons
with many friends are more likely to make new friends.  In the case of
profiles, it would mean that accounts with many profiles are more likely
to create new profiles.  Whether this is the correct explanation cannot be
explained by the data however.  Nonetheless, the distributions of pets
per household follow power laws much more closely than the number of
friends per profile. 

Thus, the account-level networks have about half the number of nodes as
the profile-level networks.  In terms of the number of edges (the volume
of the network), the account-level networks are smaller by a factor of
ten (Catster), four (Dogster) and three (Hamsterster).  The lower value
for Hamsterster can be explained by the fact that Hamsterster does not
allow friendship edges within families, but also by the fact that in
Hamsterster, the average number of friendships is lower (8.5) than
in Catster (53.3) and Dogster (37.8). 

\subsection{Are Pet Networks Scale-free?}
The distribution of the node degrees in a network is an important
characteristic of the network.  Many network models such as the
preferential attachment model \cite{b439} predict the degree
distribution to be scale-free, i.e., the number of nodes with degree $d$
to be proportional to the power $d^{-\gamma}$ for some constant
$\gamma$.  Along with estimating $\gamma$, we also used the Gini
coefficient to measure the equality of the friendship
distribution \cite{kunegis:power-law}.

\begin{figure*}
  \centering
  \subfigure[Pets per household]{
    \label{fig:pph}
    \includegraphics[width=0.9\columnwidth]{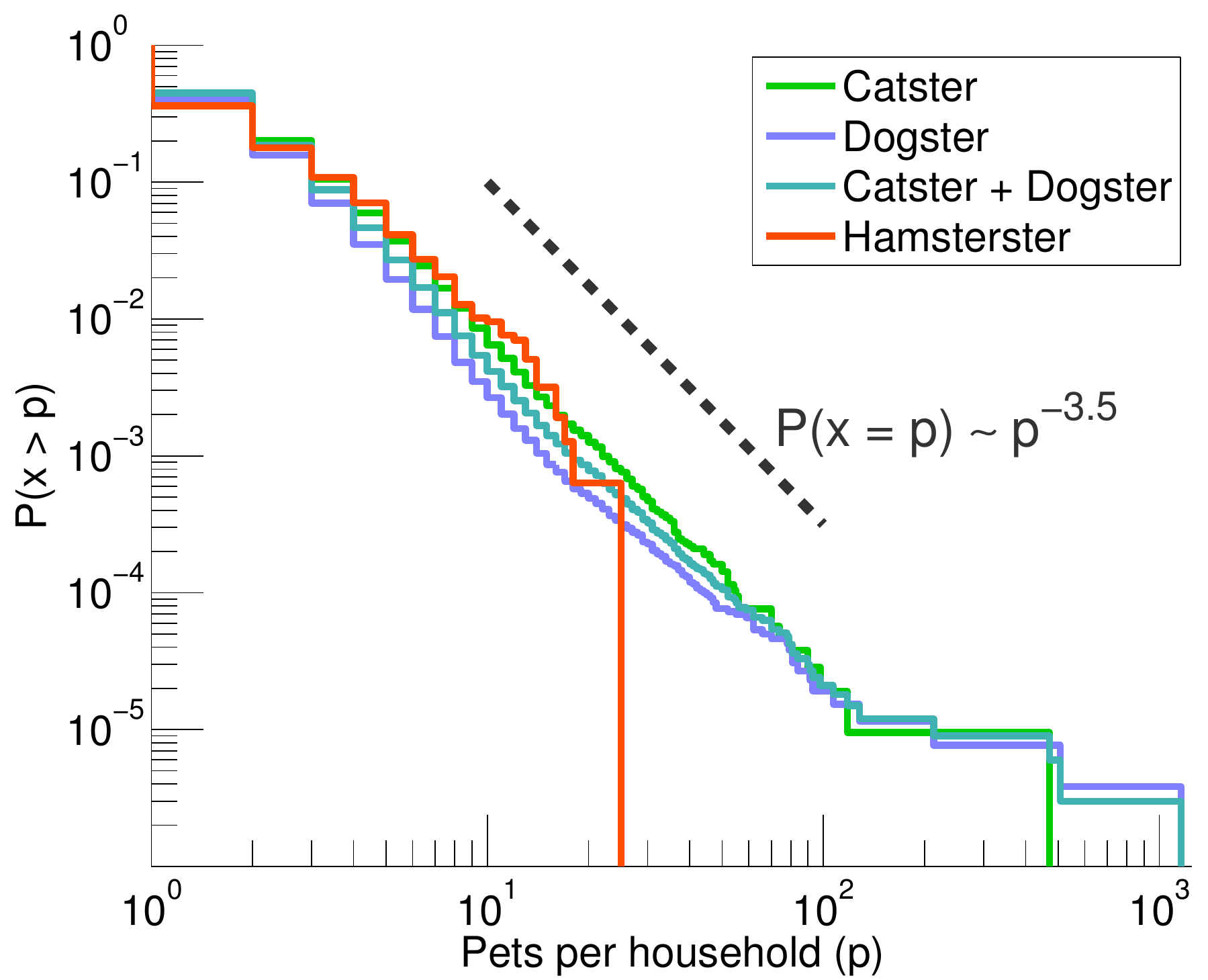}
  }
  \subfigure[Number of friendships]{
    \label{fig:degree}
    \includegraphics[width=1.0\columnwidth]{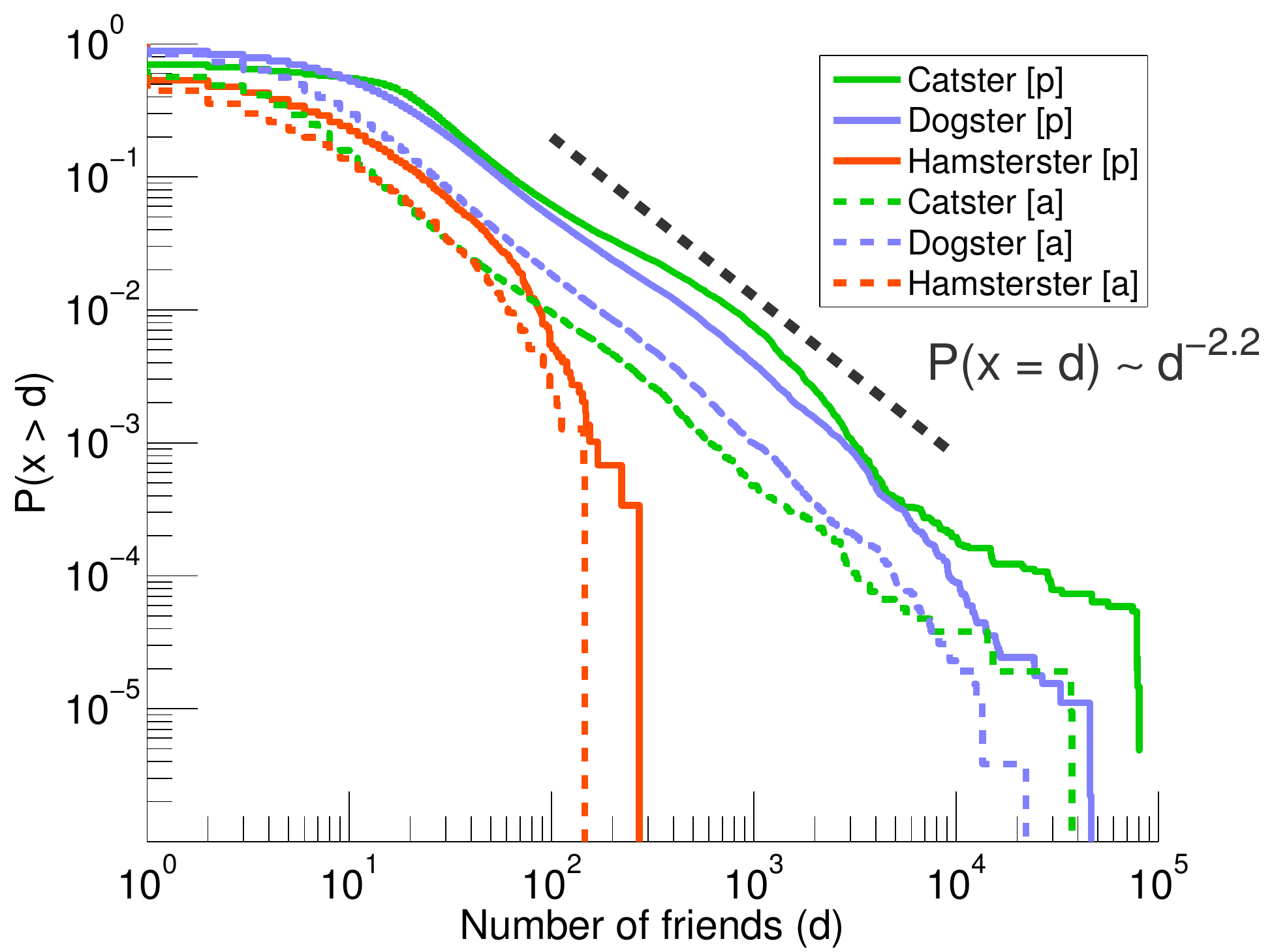}
  }
  \caption{
    Power law-like distributions in pet networks. 
    (a)~Complementary cumulative distributions of pets per household for the
    three sites, as well as for Catster and Dogster combined (as single
    accounts may create profiles on both sites). 
    (b)~Complementary cumulative degree distribution in the profile-level
    [p] and account-level [a]
    networks.   
  }
\end{figure*}

The degree distributions of the profile-level networks as well as the
account-level networks are plotted in Figure~\ref{fig:degree}, and the
values of the fitted power-law exponent $\gamma$ and the Gini
coefficient are given in Table~\ref{tab:statistic}.
The power law exponent is computed using a minimum degree $d_{\min}$,
using the robust method given in \cite[Eq.\ (5)-(6)]{b408}. 

Beyond the fact that the average degree is lower in the account-level
networks than in the profile-level networks, we observe that in Catster
and Dogster, the
power-law exponent $\gamma$ is larger for the account-level network than
for the profile-level network, while the Gini
coefficient is smaller in the account-level network than in the
profile-level network.  Both observations are consistent with each
other, as a large Gini coefficient and a small power-law exponent both
denote a more equal degree distribution \cite{kunegis:power-law}.  This
indicates that the account-level networks have a more equal distribution
of degrees than the profile-level network, i.e., the account-level
networks are more regular.  Both statistics are however
in the range usual for social networks; $\gamma$ is in the range 2.1--2.5 and the
Gini coefficient is in the range 60--70\%.

\section{Homophily in Pet Networks}
\label{sec:homophily}
The term \emph{homophily} refers to the tendency of people connected
through social ties to be similar to each other.  More precisely,
homophily can be measured by a network's assortativity with respect to a
given node property.  A network then displays positive homophily
(assortativity) when two randomly chosen connected persons are more
similar than two randomly chosen persons without regard to connections
\cite{b854}. 
Inversely, a network displays negative homophily (dissortativity) when
the opposite is the case.  
By analysing the homophily in online pet social networks, we want to
answer the following questions: 
\begin{itemize}
\item Which is higher, the homophily between friends, or within
  families?  If the homophily between friends is higher, this would
  indicate that the pets are the primary actors in the networks, and
  that families are merely organizational structures, but that a proper
  social network analysis would have to consider the pet-level network.  
  On the other hand, a higher homophily within families would indicate
  that the family (or household) is the primary social structure in the
  network, and that a social network analysis would have to consider the
  household-level structure to accurately reflect the social structure. 
\item Which profile properties correlate with two pets being friends,
  and with two pets being in the same household?  The features
  indicative of a shared household will give insight about the behavior
  of the users' choice of pets, while the features indicative of
  friendship links will be indicative of the social networking behavior
  of users. 
\end{itemize}
In order to answer these questions, we propose two complementary
assortativity coefficients that apply to multi-profile social networks,
whose ratio is measure of the relative strength of intra-household
homophily as compared to across-friendship homophily. 


\subsection{Methodology}
\label{sec:homophily:methodology}
Many different node properties
can be subject to homophily analysis, and the exact method used for
measuring it depends on the data type considered.  In the online pet
social networks we analyse, the data that can be added to a pet's
profile fall into three categories:
\begin{itemize}
  \item Categorical variables 
    \begin{itemize}
      \item The sex of a pet (male / female).  The sex is a mandatory
        field for all pets. 
      \item The race of a pet.  For cats and dogs, the race corresponds
        to the breed.  For hamsters, the race corresponds to one of
        multiple species of hamsters and gerbils.  The race is a
        mandatory field for all pets. 
      \item The pet's coloration.  The coloration is mandatory for all
        hamsters and optional for cats (69\% of profiles include it).  It
        is not used on Dogster.  
    \end{itemize}
  \item Numerical variables 
    \begin{itemize}
    \item The profile creation date.  It is known for all pets on all
      three sites. 
    \item The birth date.  The birth date is mandatory for all hamsters,
      and optional on Catster and Dogster.  It is known for 76\% of cats
      and 80\% of dogs. 
    \item The weight.  On Catster, the weight can be specified as an
      exact number in pounds, and is known for 58\% of cats.  On
      Dogster, one out of five ranges can be chosen (1--10 lbs, 11--25
      lbs, 26--50 lbs, 51--100 lbs, 100+ lbs).  The weight is not used
      on Hamsterster. 
    \item The number of friends of a pet. 
  \end{itemize}
  \item The location (``home'') of a pet can be specified on all three
    sites.  We converted the location strings to latitude-longitude
    pairs using the Google Geocoding API \cite{googlegeocoding}.  The geolocation
    is known for 68\% of cats, 78\% of dogs and 99\% of hamsters. 
\end{itemize}
We additionally use as a feature the join age, defined as the age of the
pet at the time of profile creation. 

We define two measures of assortativity for multi-profile networks:  one
that measures homophily on the
profile friendship level ($r_{\mathrm p}$)
and one that measures homophily on the account level ($r_{\mathrm a}$).  For the
friendship level, we consider the 
friendship edges between pets in the networks.  For the account level,
we consider all pairs of pets that are in the same household. 
As in most social networks, we expect to observe a certain amount of
homophily in the pet friendship network.  We further hypothesize that
the homophily between pets within a single household is larger than the
homophily for pets connected by friendship links.  Therefore, we compute
measures of homophily for both levels, based on the available pet
characteristics. 

For categorical variables, we base the assortativity coefficients on
\cite[Eq.\ (2)]{b854}.  Let $C$ be the set of possible 
values of the categorical variable, 
$P_{\mathrm x}(i,j)$
the probability that a 
randomly chosen connected pair of profiles (either via a friendship edge
for $\mathrm x = \mathrm p$, or
in the same household for $\mathrm x = \mathrm a$) are in the categories
$i\in C$ and $j\in C$ 
respectively, and $P_{\mathrm x}(i) = \sum_j P_{\mathrm x}(i,j)$.
Then, we define the friendship assortativity coefficient $r_{\mathrm p}$
and the household assortativity coefficient $r_{\mathrm a}$ using
\begin{align}
r_{\mathrm x} &= \frac {\sum_i P_{\mathrm x}(i,i) - \sum_i P_{\mathrm
    x}(i)^2} {1 - \sum_i P_{\mathrm x}(i)^2}. 
\end{align}
The assortativity coefficients defined in this way equal one for
perfect positive 
homophily, and lie between negative one and zero for negative
homophily.\footnote{$r_{\mathrm x}$ cannot be exactly $-1$; see \cite{b854} for an
  explanation.} 

For numerical variables, we use the Pearson correlation coefficient
between the numerical properties of connected pets, as defined in
\cite[Eq.\ (20)]{b854}.  Let $\mathrm{var}_{\mathrm x}(X)$ be
the variance of the numerical profile characteristic weighted by the
number of neighbors of the profile in the friendship graph, and
$\mathrm{cov}_{\mathrm x}(X, Y)$ the covariance between the 
characteristics of pairwise connected profiles, using again $\mathrm x =
\mathrm f$ for friendship connections and $\mathrm x = \mathrm a$ for
pairs of profiles of the same account.  Then the assortativity
coefficients $r_{\mathrm p}$ and $r_{\mathrm a}$ are given by
\begin{align}
  r_{\mathrm x} &= \frac {\mathrm{cov}_{\mathrm x}(X, Y)}
  {\mathrm{var}_{\mathrm x}(X)}.
\end{align}
Note that this expression is simplified from the usual Pearson
correlation coefficient because the relationships are symmetric. 
The values of $r_{\mathrm x}$ range from $-1$ to $+1$ and
are one for perfect positive homophily and $-1$ for perfect negative
homophily.  

For the geolocation, we use the distance correlation \cite{b855} as a measure of
homophily, based on the great circle distance between pairs of
locations.  Since locations are two-dimensional, the distance 
correlation is able to represent the orientation of the correlation as
does the Pearson correlation, but cannot represent the direction of the
correlation.   Therefore the distance correlation
ranges from zero to one, with one denoting perfect correlation and zero
denoting no correlation.  
The location is always the same for pet profiles created by a
single user and therefore the family-level homophily for the location is
always trivially one. 

All three types of assortativity measures are zero when neither positive nor
negative homophily is observed.  To compare the both the assortativity
coefficients on the friendship level and on the account level, we define
the multi-profile assortativity ratio of a profile characteristic as
\begin{align}
  r_{\mathrm{rel}} = \left| \frac {r_{\mathrm a}} {r_{\mathrm p}}
  \right|. 
\end{align}
By construction $r_{\mathrm{rel}}$ is larger than one if the assortativity
is higher within profiles of one account than across friendships, and
smaller than one if it is the assortativity across friendships that is
higher. 

\subsection{Discussion}

\begin{table*}
  \caption{
    \label{tab:homophily}
    Homophily analysis comparing the strength of homophily across friendships
    $r_{\mathrm p}$ and the strength of homophily within accounts
    $r_{\mathrm a}$.  The multi-profile assortativity ratio is shown
    as $r_{\mathrm{rel}}$. 
  }
  \centering
  \scalebox{0.95}{
    \begin{tabular}{ l | llr | llr | llr } 
\toprule 
& \multicolumn{3}{|c|}{\textbf{Catster}} 
& \multicolumn{3}{|c|}{\textbf{Dogster}} 
& \multicolumn{3}{|c}{\textbf{Hamsterster}} \\ 
& \textbf{\quad $r_{\mathrm p}$} & \textbf{\quad $r_{\mathrm a}$} & \textbf{$r_{\mathrm rel}$} & \textbf{\quad $r_{\mathrm p}$} & \textbf{\quad $r_{\mathrm a}$} & \textbf{$r_{\mathrm rel}$} & \textbf{\quad $r_{\mathrm p}$} & \textbf{\quad $r_{\mathrm a}$} & \textbf{$r_{\mathrm rel}$} \\ 
\midrule 
Race\textsuperscript{a} & $\phantom{-}0.0138^{++}$ & $\phantom{-}0.3137^{++}$ & 22.748 & $\phantom{-}0.1556^{++}$ & $\phantom{-}0.3065^{++}$ & 1.970 & $\phantom{-}0.0973^{++}$ & $\phantom{-}0.5349^{++}$ & 5.497 \\ 
Sex\textsuperscript{a} & $\phantom{-}0.0048^{++}$ & $\phantom{-}0.0472^{++}$ & 9.848 & $\phantom{-}0.0075^{++}$ & $\phantom{-}0.0154^{++}$ & 2.040 & $\phantom{-}0.0083^{+}$ & $\phantom{-}0.1180^{++}$ & 14.264 \\ 
Coloration\textsuperscript{a} & $\phantom{-}0.0076^{++}$ & $\phantom{-}0.0599^{++}$ & 7.864 & \qquad--- & \qquad--- & --- & $\phantom{-}0.0219^{++}$ & $\phantom{-}0.1166^{++}$ & 5.325 \\ 
Weight range\textsuperscript{a,c} & \qquad--- & \qquad--- & --- & $\phantom{-}0.1498^{++}$ & $\phantom{-}0.2590^{++}$ & 1.729 & \qquad--- & \qquad--- & --- \\ 
\#Friends\textsuperscript{b} & ${-}0.5232^{**}$ & $\phantom{-}0.7629^{**}$ & 1.458 & ${-}0.2893^{**}$ & $\phantom{-}0.6487^{**}$ & 2.242 & ${-}0.0310^{**}$ & $\phantom{-}0.6859^{**}$ & 22.108 \\ 
Birth date\textsuperscript{b} & $\phantom{-}0.0406^{**}$ & $\phantom{-}0.2930^{**}$ & 7.216 & $\phantom{-}0.0585^{**}$ & $\phantom{-}0.2114^{**}$ & 3.613 & $\phantom{-}0.3542^{**}$ & $\phantom{-}0.5614^{**}$ & 1.585 \\ 
Join date\textsuperscript{b} & $\phantom{-}0.4219^{**}$ & $\phantom{-}0.7327^{**}$ & 1.737 & $\phantom{-}0.5584^{**}$ & $\phantom{-}0.7268^{**}$ & 1.302 & $\phantom{-}0.5723^{**}$ & $\phantom{-}0.8266^{**}$ & 1.444 \\ 
Join age\textsuperscript{b} & $\phantom{-}0.0187^{**}$ & $\phantom{-}0.2600^{**}$ & 13.878 & $\phantom{-}0.0475^{**}$ & $\phantom{-}0.1738^{**}$ & 3.663 & $\phantom{-}0.0317^{**}$ & $\phantom{-}0.3615^{**}$ & 11.405 \\ 
Weight\textsuperscript{b,d} & $\phantom{-}0.0087^{**}$ & $\phantom{-}0.1827^{**}$ & 20.991 & \qquad--- & \qquad--- & --- & \qquad--- & \qquad--- & --- \\ 
Location\textsuperscript{e} & $\phantom{-}0.0888^{*}$ & \qquad--- & --- & $\phantom{-}0.1112^{**}$ & \qquad--- & --- & $\phantom{-}0.1863^{**}$ & \qquad--- & --- \\ 
\bottomrule 
\multicolumn{10}{l}{$^{++}$ and $^{+}$ denote an estimate on the error of less than 0.1\% and 1\%, respectively \cite[Eq. (5)]{b854}} \\ 
\multicolumn{10}{l}{$^{**}$ and $^{*}$ denote a $p$-value of less than 0.001 and 0.01, respectively} \\ 
\multicolumn{10}{l}{\textsuperscript{a} Categorical variable; numbers denote the assortativity coefficient \cite[Eq. (2)]{b854}} \\ 
\multicolumn{10}{l}{\textsuperscript{b} Numerical variable; numbers denote the Pearson correlation coefficient \cite[Eq. (21)]{b854}} \\ 
\multicolumn{10}{l}{\textsuperscript{c} In Dogster, the weight can only be chosen from a predefined set of ranges} \\ 
\multicolumn{10}{l}{\textsuperscript{d} In Catster, the exact pet weight can be specified} \\ 
\multicolumn{10}{l}{\textsuperscript{e} Not computed for households as all pets in one household share their location} \\ 
\end{tabular} 

  }
\end{table*}

Table~\ref{tab:homophily} shows the complete homophily analysis.  
For all features, the homophily within households is larger than the
homophily between friends, and thus all multi-profile assortativity
ratios are larger than one.  This indicates, as we would expect from
pets, that the underlying social network is primarily one of humans and
not one of pets.  However, the pet friendship network is not completely
unassortative, as it displays positive assortativity (r > 0.5) by join date
for all three sites.  

For the intra-household homophily, high values (r > 0.5) can
be observed for the join date and the number of friends. 
Small positive assortativity (r > 0.1) can be observed for the
race, the birth date, the join age, and the pet's weight. 
The largest multi-profile assortativity ratio ($r_{\mathrm{rel}} > 10$) can be
observed for the breed in Catster, the number of friends in Hamsterster,
the join age in Catster and Hamsterster, and the pet weight in Catster. 

In terms of race, Dogster has a particularly high intra-house\-hold
homophily, indicating that owners of several dogs tend to prefer dogs of
the same breed, while this is only true to a small extent for cats and
hamsters.  
The sex and coloration of pets show no homophilic tendencies.
The number of friends of a pet show negative assortativity on the
friendship, and positive assortativity within households.  This
indicates that while the friendship ties display the usual degree
dissortativity 
of real social networks, the numbers of friends of pets within one
household are similar, and therefore the degree of a pet is a function
of the owner, not of the pet. 
The homophily with respect to he join date and birth date is higher in
Hamsterster.  This is consistent with the fact that hamsters have
shorter lives. 

In conclusion, we find that the intra-household homophily is higher than
the friendship homophily.  Thus, with respect to profile features,
these pet social networks largely follow the underlying human social
networks.  This conclusion is however only based on profile properties,
and does not take into account the network structures.  Therefore, we
investigate the pet and human-level network structures in the next
section. 

\section{Relationship Between \\ Friendships and Family Ties}
\label{sec:spectral}
So far, we have analysed the friendship and family ties on an individual
level.  
We now perform several experiments to analyse the available networks as a
whole, and to determine the relationships between the friendship network
and family tie network at the structural level.  
In order to do so we extend the 
spectral diagonality test described in
\cite{kunegis:spectral-network-evolution}, which was originally used to
analyse the temporal evolution of a network, to the comparison of the
friendship network with the ownership structure in the multi-profile
network.  The result is a test that allows us to directly observe
relationships between both structures, and a measure of the consistency
between friendships and family ties. 

\subsection{Definitions}
The graphs $G_{\mathrm p}$ and $G_{\mathrm a}$ can be represented
by the adjacency matrices $\mathbf A_{\mathrm p} \in \{0,1\}^{|V|\times
  |V|}$ and $\mathbf A_{\mathrm a} \in \{0,1\}^{|W|\times |W|}$, defined
as follows:
\begin{align}
  \mathbf (A_{\mathrm p})_{uv} &= \left\{ \begin{array}{ll} 1 &
    \text{when } \{u,v\} \in V \\
    0 & \text{when } \{u,v\} \notin V \end{array} \right., \\
  \mathbf (A_{\mathrm a})_{ij} &= \left\{ \begin{array}{ll} 1 &
    \text{when } \{i,j\} \in W \\
    0 & \text{when } \{i,j\} \notin W \end{array} \right.. 
\end{align}
Both matrices are symmetric.  We also define a matrix giving the
relationship between profiles and accounts.  Let $\mathbf R \in
\{0,1\}^{|V|\times |W|}$ be the matrix defined by
\begin{align}
  \mathbf R_{ui} &= \left\{ \begin{array}{ll} 1 & \text{when } m(u) = i
    \\
    0 & \text{when } m(u) \neq i \end{array} \right..
\end{align}
$\mathbf R$ is rectangular, and by definition each row has a single
entry equaling one. 
By construction, the following relationship holds:
\begin{align}
  \mathbf A_{\mathrm a} &= [ \mathbf R^{\mathrm T} \mathbf A_{\mathrm p}
    \mathbf R ]
\end{align}
where the matrix operator $[\mathbf X]$ rounds all nonzero entries of
$\mathbf X$ to one, and all diagonal entries to zero.  We also define
the family matrix $\mathbf F \in \{0,1\}^{|V|\times |V|}$ whose entries
equal one when two profiles are managed by the same account and zero
otherwise:
\begin{align}
  \mathbf F_{uv} &= \left\{ \begin{array}{ll} 1 & \text{when } m(u) = m(v)
    \\
    0 & \text{when } m(u) \neq m(v) \end{array} \right.
\end{align}
The following relationship can then be established:
\begin{align}
  \mathbf F &= \mathbf R \mathbf R^{\mathrm T}
\end{align}
Note that the diagonal elements of $\mathbf F$ are all one, since every
profile is in the same account as itself. 

\subsection{Methodology}
We seek to compare the friendship-level network and the family tie
network using a spectral diagonality test, a technique that was
initially introduced to study time-evolving networks under the spectral
evolution hypothesis, i.e., the hypothesis that under time evolution,
the eigenvalues of a network's adjacency matrix change while its
eigenvector stay nearly constant
\cite{kunegis:spectral-network-evolution}.  Two matrices with the same
eigenvectors are related by spectral transformations
\cite{kunegis:spectral-transformation}, and if they are adjacency
matrices their relationship indicates how the one type of edge is
related to the other type of edge.  If $\mathbf A_1$ and $\mathbf A_2$
are the adjacency matrix of a single network at two different
timepoints and defined on the same node set, then the spectral diagonality
test first computes the rank-$k$ eigenvalue decomposition
\begin{align}
  \mathbf A_1 &= \mathbf U \mathbf \Lambda \mathbf U^{\mathrm T},
\end{align}
and then sets out the write an eigenvalue decomposition-like expression
for $\mathbf A_2$, using the same eigenvector matrix $\mathbf U$ as for
the first matrix:
\begin{align}
  \mathbf A_2 &= \mathbf U \mathbf \Delta \mathbf U^{\mathrm T}
\end{align}
If both $\mathbf A_1$ and $\mathbf A_2$ have the same set of
eigenvectors, then the last equation is a proper rank-$k$ eigenvalue
decomposition of $\mathbf A_2$, and $\mathbf \Delta$ gives its
eigenvalues. Solving for $\mathbf \Delta$ gives
\begin{align}
  \mathbf \Delta &= \mathbf U^{\mathrm T} \mathbf A_2 \mathbf U. 
\end{align}
If the $k$-by-$k$ matrix $\mathbf \Delta$ is diagonal, then the spectral
evolution hypothesis is true, and if $\mathbf \Delta$ is nearly
diagonal, then the hypothesis is nearly true.  Furthermore, comparing
the diagonal entries of $\mathbf \Lambda$ and $\mathbf \Delta$
gives an indication as to the actual algebraic function connecting the two
matrices, such as matrix powers or
exponentials~\cite{kunegis:spectral-transformation}. 

\begin{figure*}
  \centering
  \subfigure[Catster]{\includegraphics[height=0.34\columnwidth]{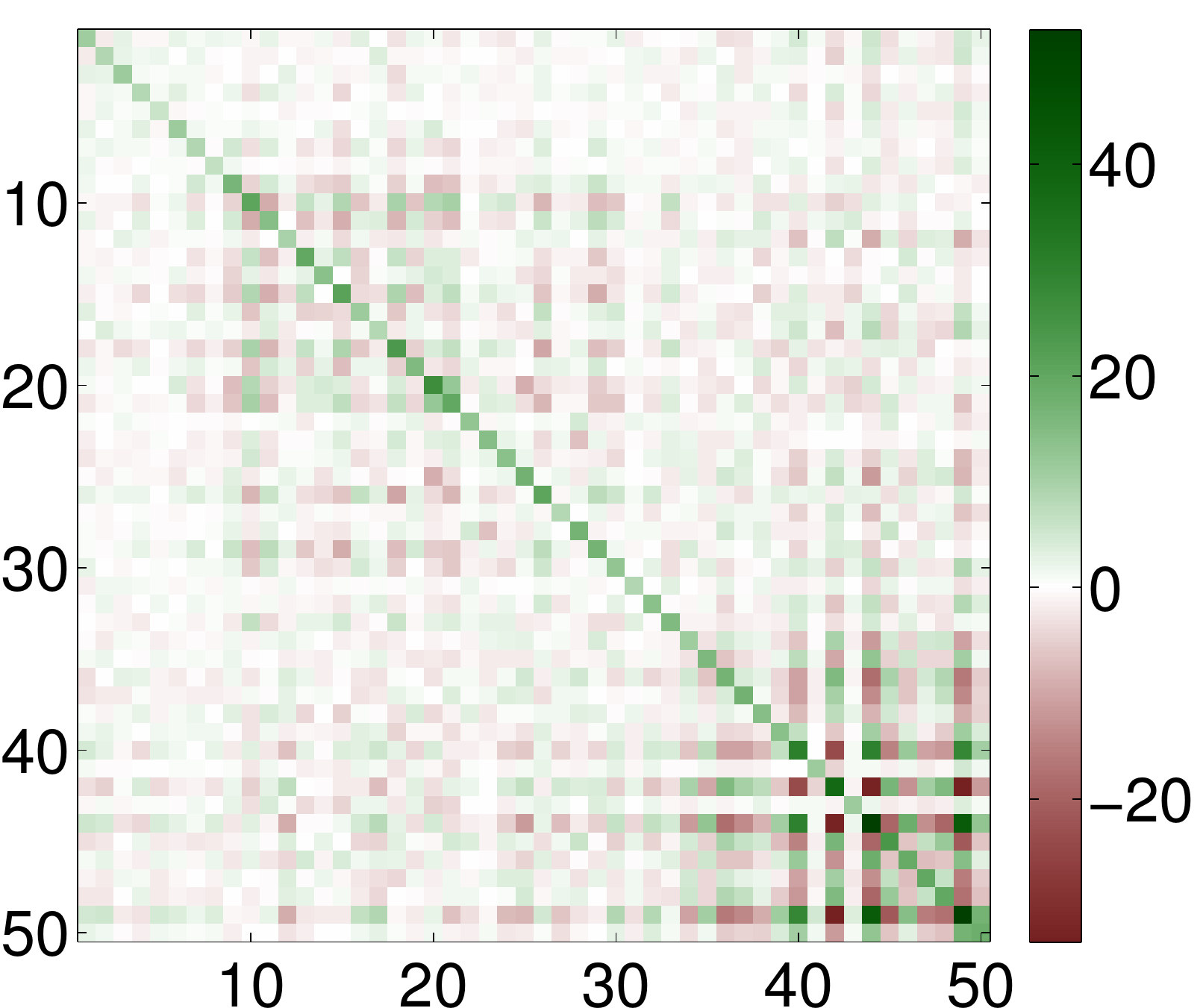}}
  \subfigure[Dogster]{\includegraphics[height=0.34\columnwidth]{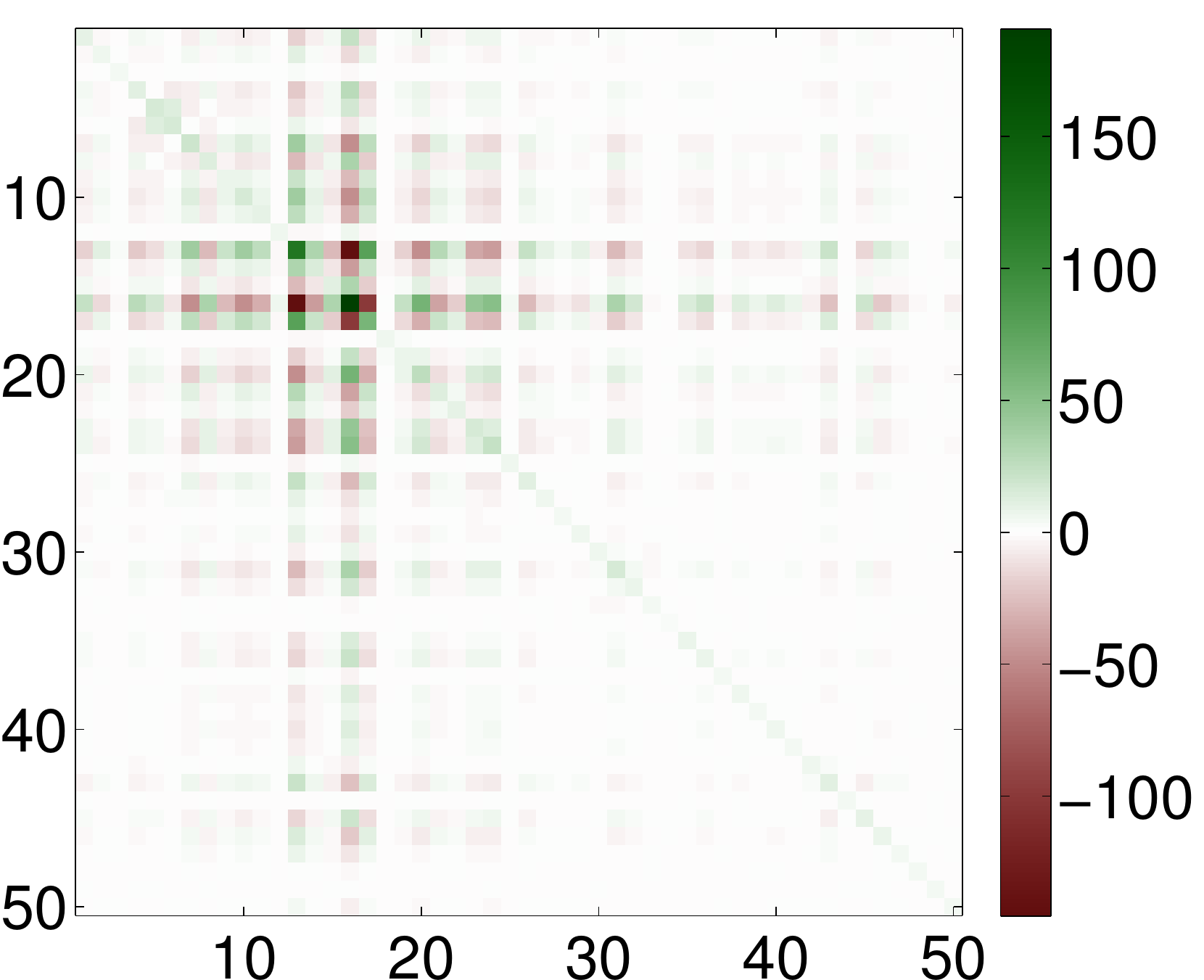}}
  \subfigure[Hamsterster]{\includegraphics[height=0.34\columnwidth]{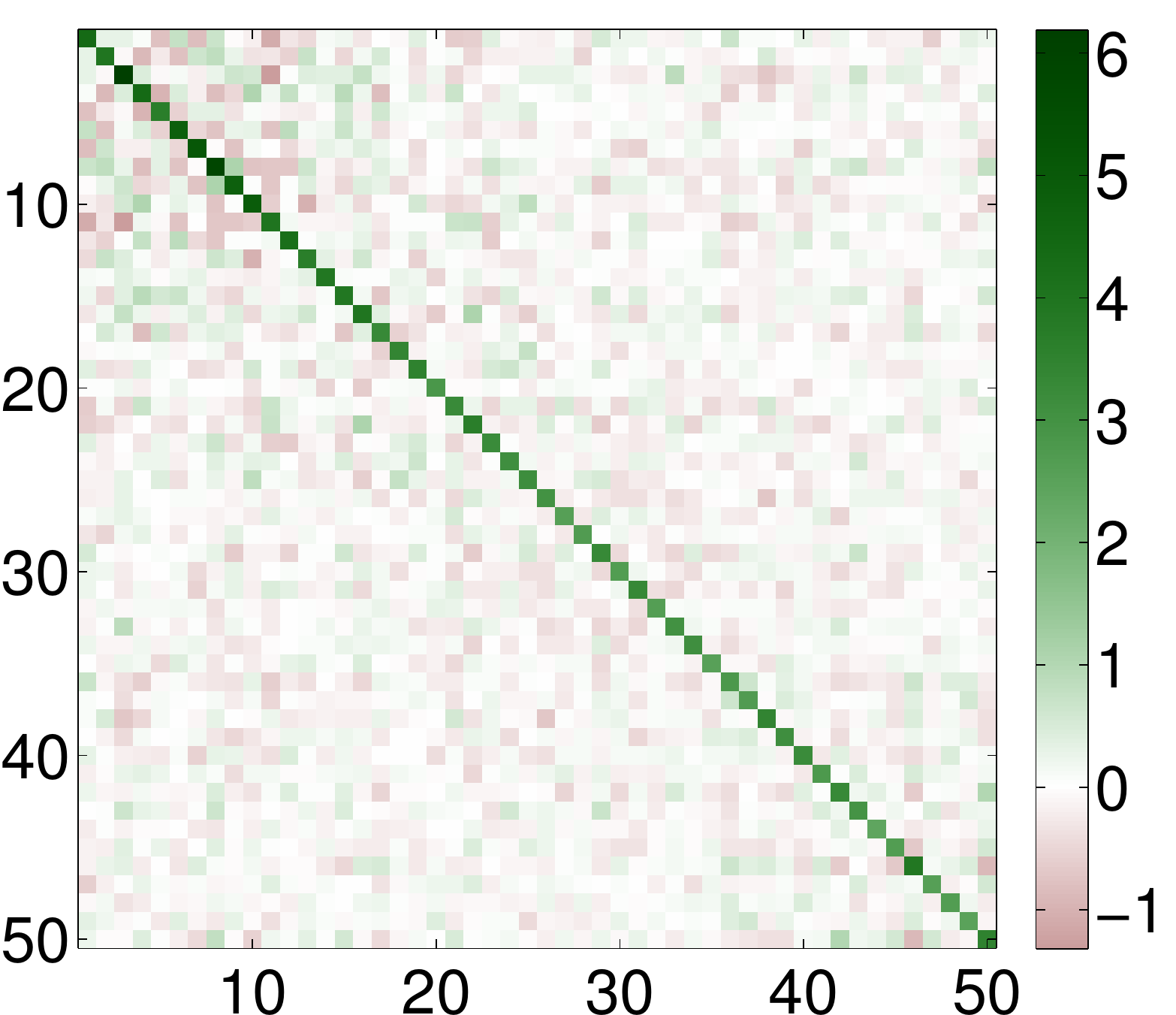}}
  \\
  \subfigure[Catster]{\includegraphics[height=0.36\columnwidth]{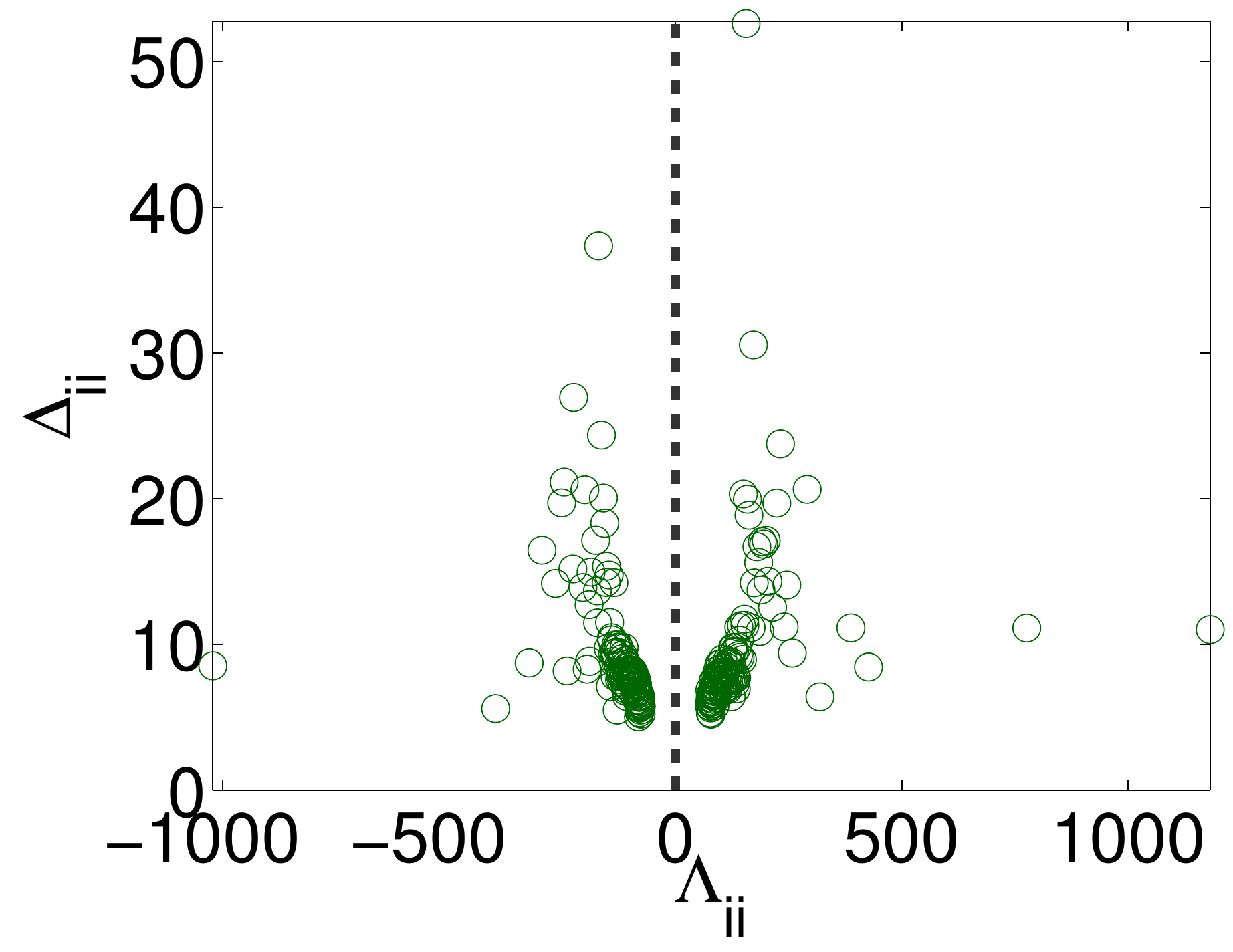}}
  \subfigure[Dogster]{\includegraphics[height=0.36\columnwidth]{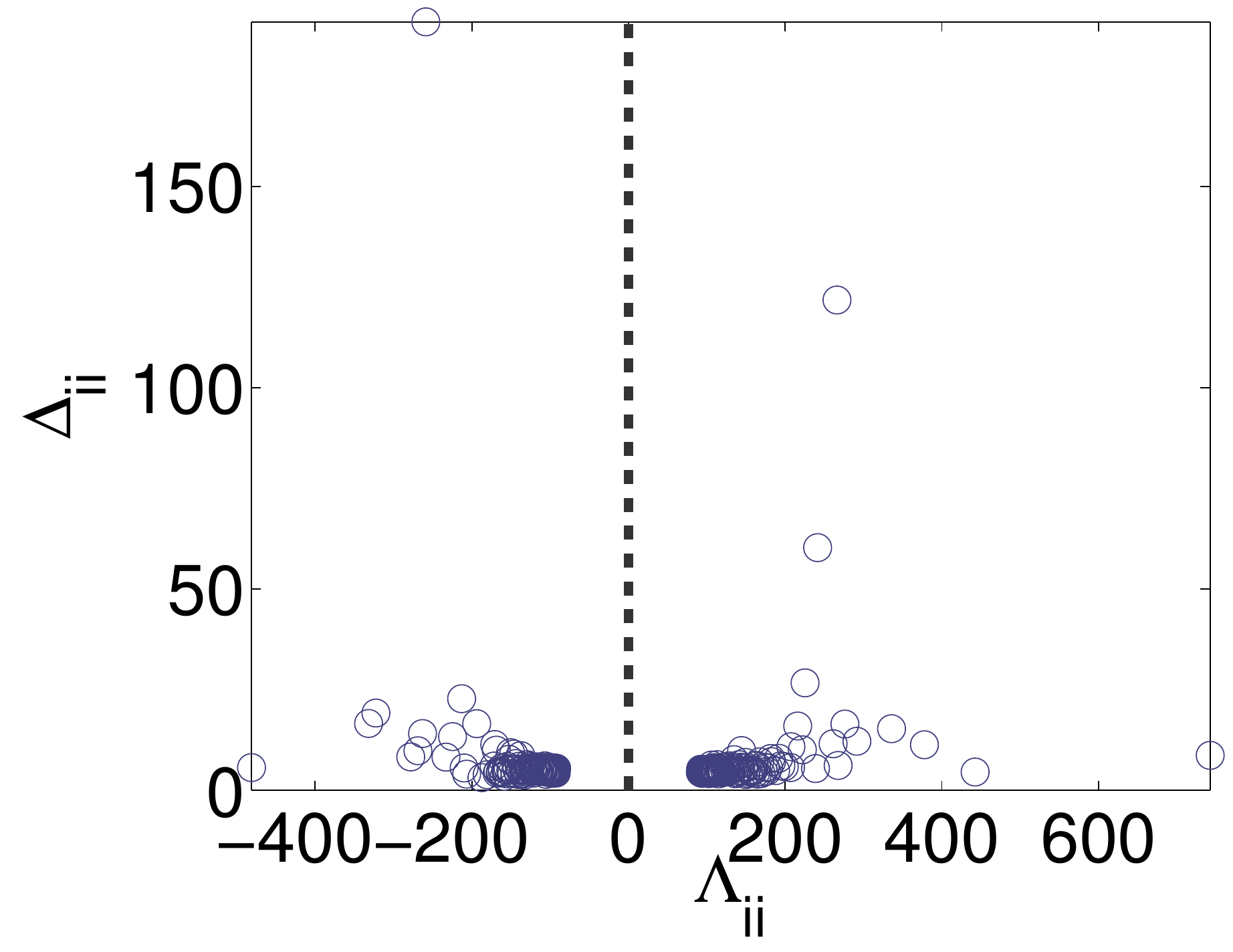}}
  \subfigure[Hamsterster]{\includegraphics[height=0.36\columnwidth]{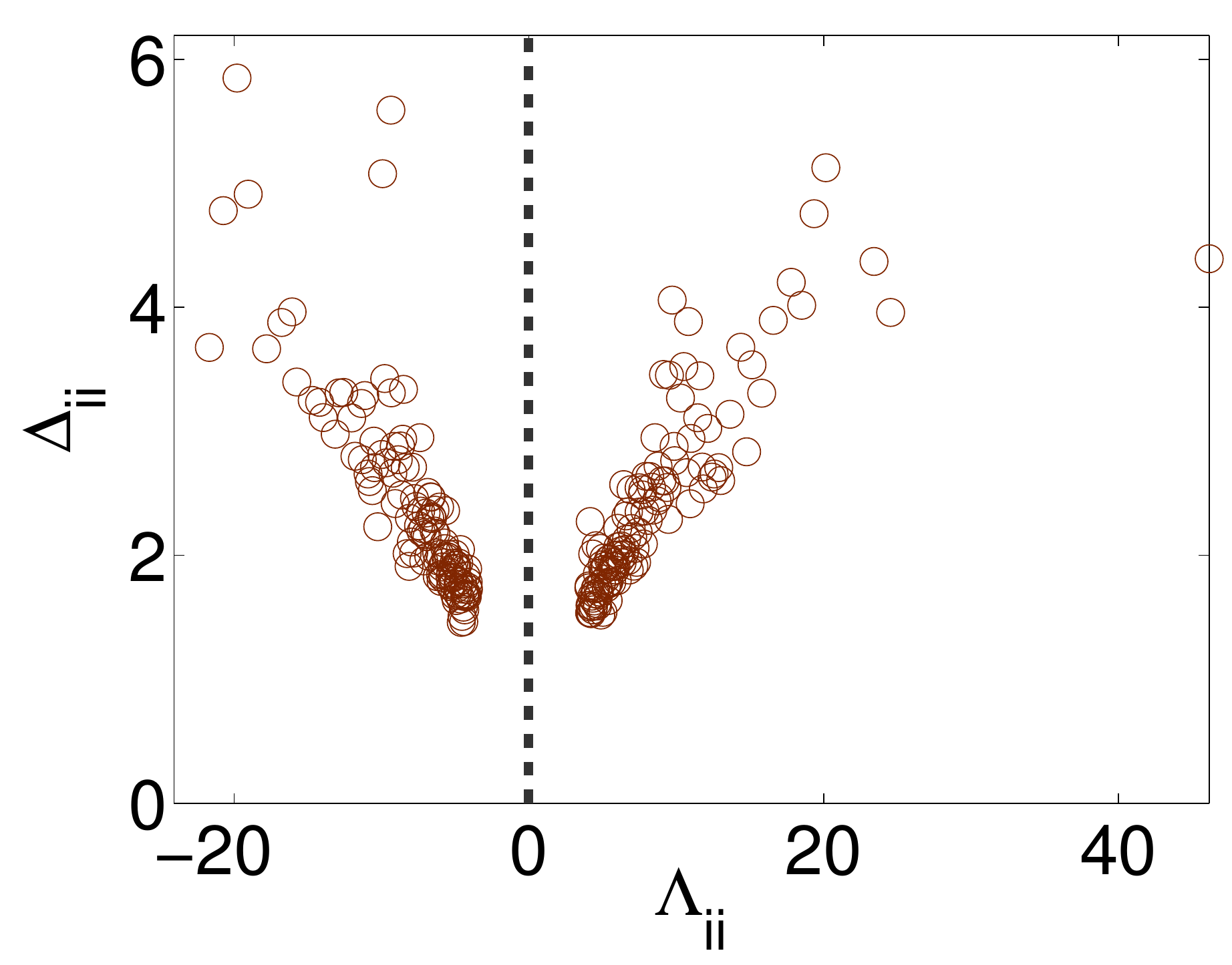}}
  \caption{
    \label{fig:delta}
    The spectral diagonality test matrices $\Delta$ for the three
    sites, restricted to the topmost $50\times 50$ submatrix
    corresponding to largest eigenvalues of $\mathbf \Lambda$. 
    (a-c)~The spectral diagonality test matrix $\mathbf \Delta$.
    (d-f)~Comparison plot between the diagonal entries of $\mathbf
    \Lambda$ and $\mathbf \Delta$. 
  }
\end{figure*}

In the context of multi-profile networks, our goal is to learn the
relationship between the friendship network and the
family relationships.  Thus, we apply the spectral diagonality test to
the matrices $\mathbf A_{\mathrm p}$ and $\mathbf F$.  First, we compute
the rank-$k$ eigenvalue decomposition of the friendship adjacency matrix:
\begin{align}
  \mathbf A_{\mathrm p} &= \mathbf U \mathbf \Lambda \mathbf U^{\mathrm T}
\end{align}
We then compute $\mathbf \Delta$:
\begin{align}
  \mathbf \Delta &= \mathbf U^{\mathrm T} \mathbf F \mathbf U \nonumber 
  = \mathbf U^{\mathrm T} \mathbf R \mathbf R^{\mathrm T} \mathbf U
\end{align}
Testing the $k$-by-$k$ matrix $\mathbf \Delta$ for diagonality then
gives an indication whether both matrices are related, and the
relationship between the matrices $\mathbf \Lambda$ and $\mathbf \Delta$
gives an indication of the path relationships between friend and family
relations. 
We use the value $k=250$ in all calculations. 
We additionally also define the coefficient of diagonality, which
measures what proportion of the matrix $\mathbf F$ is explained by a
spectral transformation of $\mathbf A_{\mathrm p}$.  We define the
coefficient of diagonality as
the proportion of square entry weights in $\mathbf \Delta$ that lie on
the diagonal:
\begin{align}
  \delta &= \frac{\sum_i \mathbf \Delta_{ii}^2}{\sum_{i,j} \mathbf \Delta_{ij}^2}
\end{align}
The coefficient ranges from zero to one, and attains one when the two
matrices have the exact same eigenvectors.  
The denominator is the squared Frobenius norm of $\mathbf \Delta$,
and since the Frobenius norm is invariant under orthogonal
transformations, it follows that $\delta$ is the largest number such
that $\mathbf F$ can be written as a sum of a spectral transformation of
$\mathbf A_{\mathrm p}$ and another matrix.  Thus, $\delta$ denotes to
what extent the family relationships are represented by friendships. 
Note that this coefficient works in
an opposite way to well-known co-spectrality measures \cite{b858}, which aim to
measure how similar the eigenvalues of two matrices are, while $\delta$
aims to measure to what extent they share the same eigenvectors. 

\subsection{Experiments}
We compute the matrix $\mathbf \Delta$ as described above for the three sites,
and show the result in Figure~\ref{fig:delta} (a-c).  Furthermore,
Table~\ref{tab:diagonality} shows the diagonality coefficient $\delta$ of the
tests. 
The results show that all three datasets display a partial diagonality for
the matrix $\mathbf \Delta$.  The diagonality coefficient $\delta$ is
20\% for 
Dogster, 28\% for Catster, and 55\% for Hamsterster.  We may conclude
from this that friendship links and family ties are the most consistent
with each other on Hamsterster.  All three results are consistent with
temporal network evolution results given in
\cite{kunegis:spectral-network-evolution}. 

Additionally, we show in Figure~\ref{fig:delta} (d-f) the relationship
between the diagonal elements of the matrix $\mathbf \Lambda$ (the
eigenvalues of $\mathbf A_{\mathrm p}$) and the diagonal elements of
$\mathbf \Delta$.  This type of plot serves to find out which matrix
functions best maps one matrix to another
\cite{kunegis:spectral-transformation}.  The three mappings seen in the
plot allow us to draw two conclusions.  First, the plots are nearly
symmetrical around the Y axis, indicating that the best mapping matrix
function is an even function, i.e., paths of even lengths of friendships
should be used to predict family ties.  Secondly, for Catster and
Hamsterster, the distribution of eigenvalues follows a nearly
linear trend, indicating that a linear spectral graph transformation may
be used, i.e., only short paths are relevant, and longer even paths (of
length four, six, etc.) are not relevant.  This is however not observed
for Dogster.

\begin{table}
  \caption{
    \label{tab:diagonality}
    The diagonality coefficient $\delta$ for the three sites. 
  }
  \centering
  \begin{tabular}{ l r }
    \toprule
    \textbf{Dataset} & \textbf{$\delta$} \\
    \midrule
    Catster     & 0.2754 \\
    Dogster     & 0.2013 \\
    Hamsterster & 0.5512 \\
    \bottomrule
  \end{tabular}
\end{table}

\section{Predicting Family Ties}
\label{sec:prediction}
A family tie can be thought to exist between two pets that are in the
same family, i.e., whose profiles were created by the same user
account.  While on Catster, Dogster and Hamsterster tie information is
readily available under the ``Meet My Family'' header, the fact that two
profiles were created by the same person cannot be easily verified on
other online social networking platforms.  Therefore, pet networks
present an opportunity to study the prediction problem of detecting
which profiles were created by the same account, since they provide
complete ground truth data for an evaluation of the task.  Thus, 
we analyse in this section the task of predicting that two pets are in
the same family, 
given only friendship links and pet-level profile metadata.  This
allows us to determine how well it 
can be predicted whether two profiles are from the same account, even
when that information is not public.  
Since we have multiple types of profile
data available, we can investigate which profile data allows to do this
how well.  Also, the experiment serves to find out which properties of
pets are consistent within a household, and which are independent
of a household. 

\subsection{Prediction Methods}
Given a multi-profile social network $G=(V,W,E,m)$, we want to predict whether
two profiles are managed by the same account, i.e., information
contained in $W$ and $m$, using only the profile-level network
$G_{\mathrm p}=(V,E)$, including the metadata associated with it.  In
the case of pet social networks, we use the available pet profile
information along with the pet-level friendship links for learning. 
We investigate the following indicators (i.e., features), each of which
applies to a pair of profiles $\{u,v\}$:
\begin{itemize}
  \item Degree difference:  The difference of degrees. 
  \item Friend:  This feature is one if there is a friendship between
    the two profiles and zero otherwise. 
  \item Common friends:  The number of common friends between the two profiles. 
  \item Jaccard index:  The Jaccard index between the sets of friends of the
    two profiles \cite{b857}.  This is related to the number of common friends,
    being normalized by the number of friends of either profile. 
  \item Same race, sex, coloration, location, join date and weight:
    These features are one if the corresponding profile information is
    equal, and zero otherwise.
  \item Birth date, join date, join age and weight difference:  The
    negative 
    absolute difference between the corresponding values for the two
    profiles.  We take the negative since we expect a small difference
    to be indicative of a same household, due to intra-household
    homophily.  
\end{itemize}
The exact definitions are given in Table~\ref{tab:def}. 
We do not use geographical distance between the two profiles, because we
know that if the distance is larger than zero, then the profiles must be
in distinct households.  Thus, we only the the ``same location''
feature.  Note also that the geolocation is given only up to the city
level, i.e.\ all pets in New York City will be counted as having the
same location, leading to a large number of pets from different
households but with the exact same location. 

\begin{table}
  \caption{
    \label{tab:def}
    Definitions of the features used for family tie prediction.  
    Each feature is given as a function of an unordered profile pair
    $\{u,v\}$. 
  }
  {\centering
    \begin{tabular}{ l | l }
    \toprule
    \textbf{Feature} & \textbf{Definition} \\
    \midrule
    Degree difference\textsuperscript{a} &
    $|\log(1 + d(u)) - \log(1 + d(v))|$ \\
    Friend & 
    $\left\{ \begin{array}{ll} 1 & \text{when } \{u,v\} \in E \\
      0 & \text{otherwise} \end{array} \right.$ \\
    Common friends & 
    $\left|\left\{ w \in V \mid \{u,w\},
    \{v,w\} \in E \right\}\right|$ \\
    Jaccard index &
    \hspace{-0.23cm}\scalebox{1.45}{
    $\frac
    {\left|\left\{ w \in V \mid \{u,w\} \in E \wedge \{v,w\} \in E \right\}\right|}
    {\left|\left\{ w \in V \mid \{u,w\} \in E \vee \{v,w\} \in E \right\}\right|}
    $} \\
    Same $X$ & 
    $\left\{ \begin{array}{ll} 1 & \text{when } X(u) = X(v) \\
      0 & \text{otherwise} \end{array}\right.$ \\
    Difference in $X$ &
    $-|X(u) - X(v)|$ \\
    \bottomrule
  \end{tabular} }
  \textsuperscript{a}
  We use the logarithm because the distribution of 
  degrees is better distributed on a logarithmic scale.  
  The additive term of one is used to take into account
  degrees of zero. 
\end{table}

We also perform a logistic regression prediction, combining all features
given above.  Let $f_i(u,v)$ be the values for all features $i$
enumerated above.  Then, a logistic regression model takes the form
\begin{align}
  f_{\mathrm{reg}}(u,v) &= \left( 1 + \exp\left\{-a - \sum_i b_i
  f_i(u,v)\right\} \right)^{-1}. 
\end{align}
The regression parameters $b_i$ as well as $a$ are learned using a
training set of profile pairs.  The training profile pairs are sampled
from each dataset such that it contains $e$ pairs of profiles that are
in the same household and $e$ pairs of profiles that are not in the same
profile.  This training set is disjoint from the test set defined in a
similar way below. 

\subsection{Experimental Setup}
In order to measure the accuracy of each prediction method, we use a 
test set defined in the same manner as the training set, i.e.,
we randomly sample $e$ pet pairs known to be in the same
family, and $e$ pet pairs known not to be in the same family.  
This test set is disjoint from the training set used for learning the
regression parameters. 
The accuracy of the prediction methods is measured using the area under
the curve (AUC) \cite{b366}, which measures the probability that our
prediction gives the correct ordering when applied to two randomly
chosen pairs of profiles.  Thus, the AUC is 1/2 for a random prediction,
and one for a perfectly accurate prediction.  It is less than 1/2 for
inverted predictions, i.e.\ predictions methods that become better when
their values are negated.  A perfectly inaccurate prediction has an AUC of zero. 
Table~\ref{tab:features} gives the AUC values for each method
separately and for the regression predictions, as well as the learned
regression weights for each of the three sites. 

\begin{table*}
  \caption{
    \label{tab:features}
    Results of family tie prediction. 
  }
  \centering
  \begin{tabular}{ l | r r r | r r r }
\toprule
 & \multicolumn{3}{|c|}{\textbf{AUC}} & \multicolumn{3}{|c}{\textbf{Regression weights}} \\ 
\textbf{Feature} & \textbf{Cat} & \textbf{Dog} & \textbf{Ham.} & \textbf{Cat} & \textbf{Dog} & \textbf{Ham.} \\ 
\midrule
Degree difference  & $82.3\%$  & $75.7\%$  & $72.3\%$  & $0.09$  & $-0.27$  & $0.22$ \\ 
Friend\textsuperscript{a}  & $50.3\%$  & $50.6\%$  & ---  & $4.83$  & $3.76$  & --- \\ 
Common friends  & $79.0\%$  & $91.5\%$  & $71.7\%$  & $-0.46$  & $0.71$  & $4.98$ \\ 
Jaccard index  & $82.8\%$  & $92.2\%$  & $76.2\%$  & $5.78$  & $9.73$  & $1.25$ \\ 
Same race  & $66.4\%$  & $66.2\%$  & $76.4\%$  & $1.32$  & $3.08$  & $0.92$ \\ 
Same sex  & $51.9\%$  & $50.3\%$  & $54.2\%$  & $0.07$  & $0.02$  & $-0.09$ \\ 
Same coloration\textsuperscript{b}  & $57.2\%$  & ---  & $59.4\%$  & $0.95$  & ---  & $5.59$ \\ 
Same location  & $87.2\%$  & $90.3\%$  & $99.6\%$  & $11.02$  & $8.92$  & $21.21$ \\ 
Birth date difference  & $53.7\%$  & $50.1\%$  & $73.5\%$  & $-0.41$  & $-0.30$  & $0.42$ \\ 
Same join date  & $79.7\%$  & $74.6\%$  & $78.2\%$  & $6.08$  & $5.44$  & $6.21$ \\ 
Join date difference  & $90.8\%$  & $87.6\%$  & $91.9\%$  & $1.19$  & $0.87$  & $-0.24$ \\ 
Join age difference  & $52.7\%$  & $48.7\%$  & $66.2\%$  & $0.42$  & $0.30$  & $-0.88$ \\ 
Weight difference\textsuperscript{c}  & $41.6\%$  & ---  & ---  & $-0.01$  & ---  & --- \\ 
Same weight\textsuperscript{c}  & ---  & $61.9\%$  & ---  & ---  & $0.52$  & --- \\ 
\midrule
Regression  & $99.3\%$  & $99.6\%$  & $99.9\%$ \\ 
\bottomrule
\multicolumn{7}{l}{\textsuperscript{a} Hamsterster does not allow friendship links within one household. } \\ 
\multicolumn{7}{l}{\textsuperscript{b} Dogster does not allow to specify a dog's coloration.} \\ 
\multicolumn{7}{l}{\textsuperscript{c} Catster allows exact weights and Dogster has weight ranges.} \\ 
\end{tabular}

\end{table*}

\subsection{Discussion}
We observe that in all three sites, pets in the same household can be
detected with an AUC of over 99\% using the regression predictor.  This
means that given two pairs of 
pets, one of which from the same household and one of which from two
different households, our algorithm will detect which is which in over
99\% of cases.  This high value can be explained by the fact that
certain individual indicators are already highly indicative of family
ties.  

The best individual predictor, the join date difference, achieves an AUC
near to 
90\% for all three sites, indicating that users often create multiple
pet accounts in quick succession.  This may be explained by the fact
that the sites have only been in operation for a decade.  After a longer
time period of observation, we may expect this number to go down.  In
contrast to this, the birth date of a pet is not a good indicator for
being in the same household (AUC near to 50\% for Catster and Dogster),
indicating that users of the pet social networks do not have pets all
born in quick succession; this is consistent with the behavior of many
people acquiring new pets only after old ones die.  

The location is a good individual indicator too, as by construction pets
of the same household must have the same location. 

Properties of pets such as the sex, the race, the coloration and the weight
are not good indicators, with most AUC values not differing much from 1/2.
The highest AUC values among
these is achieved by the species of hamsters (76\%), the breed of cats
and dogs (66\%) and the weight ranges on Dogster (62\%).  This indicates
that there is a slight tendency for owners to own pets of the same
breed, and dogs of comparable weight.  The failure of cat weight's to
predict anything can be explained by the low variance in cat weights in
general, as compared to the high variance of dog weights. 

The indicators based on the friendship network achieve AUC values from
70\% to 90\%, also indicating good prediction performance.  The only
exception is the existance of a friendship link itself, whose AUC is very
near to 1/2.  We may interpret this as users not being sure what to make
of the possibility to connect two of their own pets with a friendship
link; some users do it and some do not.  This result is consistent with
the symmetric shape of the plots in Figure~\ref{fig:delta} (d-f), which
indicate that paths of even length of friendship links should be
used. 

\section{Summary and Conclusions}
\label{sec:conclusion}
We have analysed the three online pet social networks Catster, Dogster and
Hamsterster under the aspect of them being multi-profile networks, as
they allow individual users to create any number of profiles, for each
of their pets.  We 
have shown that multi-profile networks can be analysed on
two levels:  the profile level and the account level.  Our experiments
showed that the two networks are related, but not identical, as the
profile-level network is smaller, has smaller degrees, has a more equal
degree distribution, less clustering and lower average path lengths.  
We also showed that a multi-profile network implicitly contains household
links, and therefore a comparison between friendship and household links
can be performed.  We confirmed through a homophily analysis that
intra-household homophily is higher than across-friendship homophily,
and defined the multi-profile assortativity ratio in order to measure
that difference.  
In experiments, we found that the pet breed, join age
and weight display the highest differences.  
Through extended spectral tests of diagonality, we were able to discover
the relationship between friendships links and family ties in the
network. 
Finally, we showed that it is possible to predict whether two profiles
were created by the same user with a very high precision.  
In regards to this high precision, we conclude that it should be
possible in principle to analyse
the behavior of users creating multiple accounts on social networking
platforms where this is not allowed.  While corresponding datasets are
inherently difficult to come by, a corresponding analysis would shed
light on user behavior in terms of whether the profiles they create
can be considered individual actors in the social network, or whether
the person-level network should rather be considered.  Although the methods
developed in this paper can be applied to such datasets, we do not
expect the individual numerical results to hold for the individual
features, as users knowing that 
the creation of multiple accounts is not allowed can be expected to
behave in a largely different way than users who are allowed to do
this. 

\bibliographystyle{abbrv}
\bibliography{petster,ref,kunegis,konect}

\end{document}